\documentclass[a4paper,twocolumn,11pt,accepted=2021-06-21]{quantumarticle}
\pdfoutput=1
\usepackage[utf8]{inputenc}
\usepackage{graphicx}
\usepackage[dvipsnames]{xcolor}
\usepackage{amssymb,amsmath}
\usepackage[numbers,sort&compress]{natbib}

\usepackage{mathbbol}

\begin{document}

\title{Probing the edge between integrability and quantum chaos in interacting few-atom systems}

\author{Thom\'as Fogarty}
\affiliation{Quantum Systems Unit, Okinawa Institute of Science and Technology Graduate University, Onna, Okinawa 904-0495, Japan}
\author{Miguel \'Angel Garc\'{\i}a-March}
\affiliation{Instituto Universitario de Matem\'{a}tica Pura y Aplicada, Universitat Polit\`ecnica de Val\`encia, E-46022 Val\`encia, Spain}
\author{Lea F. Santos}
\affiliation{Department of Physics, Yeshiva University, New York, New York 10016, USA}
\author{N.L. Harshman}
\affiliation{Department of Physics, American University, 4400 Massachusetts Ave. NW, Washington, DC 20016, USA}

\maketitle
\begin{abstract}
	Interacting quantum systems in the chaotic domain are at the core of various ongoing studies of many-body physics,  ranging from the scrambling of quantum information to the onset of thermalization. We propose a minimum model for chaos that can be experimentally realized with cold atoms trapped in one-dimensional multi-well potentials. We explore the emergence of chaos as the number of particles is increased, starting with as few as two, and as the number of wells is increased, ranging from a double well to a multi-well Kronig-Penney-like system.  In this way, we illuminate the narrow boundary between integrability and chaos in a highly tunable few-body system. We show that the competition between the particle interactions and the periodic structure of the confining potential reveals subtle indications of quantum chaos for 3 particles, while for 4 particles stronger signatures are seen. The analysis is performed for bosonic particles and could also be extended to distinguishable fermions.
\end{abstract}	


\section{Introduction}

The interest in quantum chaos, especially when caused by the interactions between particles, has grown significantly in the last few years due to its relationship with several questions of current experimental and theoretical research that arise in atomic, molecular, optical, condensed matter, and high energy physics, as well as in quantum information science. In interacting many-body quantum systems, quantum chaos ensures thermalization and the validity of the eigenstate thermalization hypothesis (ETH) \cite{Zelevinsky1996,Borgonovi2016,Dalessio2016}, hinders localization~\cite{Schreiber2015,Nandkishore2015,Luitz2017}, facilitates the phenomenon of many-body quantum scarring~\cite{Bernien2017,Turner2018}, leads to diffusive transport~\cite{BertiniARXIV,Znidaric2020} and causes the fast spread of quantum information~\cite{Roberts2015,Maldacena2016JHEP}.

Quantum chaos refers to properties of the spectrum and eigenstates that appear in the quantum domain when the classical counterpart of the system is chaotic in the sense of mixing and positive Lyapunov exponent. The features are similar to what we find in random matrix theory~\cite{MehtaBook}, namely the eigenvalues are strongly correlated~\cite{Guhr1998} and the eigenstates in the mean-field basis are close to random vectors~\cite{Zelevinsky1996}. This quantum-classical correspondence is well established for systems with few degrees of freedom~\cite{StockmannBook}, such as billiards, the kicked rotor, and the Dicke model, where the source of chaos is respectively the shape of the billiard, the strength of the kicks, and the collective interaction between light and matter. In the case of systems with many interacting particles, the semiclassical analysis is challenging and sometimes not well defined, so the common approach has been to refer to many-body quantum systems that present the above mentioned properties of spectrum and eigenstates as chaotic, even when the classical limit is not analyzed. 

The purpose of this work is to identify a minimum model of interacting particles that is chaotic and that can be experimentally studied with cold atoms. There are theoretical examples in the literature of quantum systems with only 3 or 4 interacting particles that already exhibit chaotic properties. They include the cesium atom, which has 4 valence electrons~\cite{Flambaum1994}; systems composed of 4 particles of unequal masses in a harmonic trap~\cite{Harshman2017} and 3 particles with unequal masses on a ring~\cite{huber_morphology_2021}; 3 or 4 excitations in spin-1/2 chains with short-range~\cite{Schiulaz2018} or long-range couplings~\cite{Zisling2021}; and even spin-1/2 chains with only 3 sites~\cite{Mirkin2021}. In the context of thermalization due to chaos, we also find works that obtained the Fermi-Dirac distribution in systems with only 4 particles~\cite{Schnack1996,Schnack2000,Flambaum1996,Flambaum1997,Izrailev2001}. 

We consider a one-dimensional system with $N$ identical particles that is split into wells separated by delta-function barriers. A single barrier defines a double-well system and many barriers results in the finite Kronig–Penney model \cite{Zoller2016,Reshodko_2019}. We focus on the sector of states symmetric under particle exchange and spatial parity, and that the particles interact via contact interactions which are modeled with a delta function as in the Lieb-Liniger Hamiltonian \cite{LLmodel1963,1998Olshanii}. The system is integrable when there are finite barriers \emph{or} finite interactions (not both). It is also  solvable in the limiting cases of infinite barrier strength and infinite interaction strength. However, we provide numerical evidence that when the interaction strength and the barrier strength are simultaneously finite, integrability is broken. In this case, strong signatures of quantum chaos emerge for $N = 4$ particles in the presence of just one barrier (double-well system). We also demonstrate that the signatures of quantum chaos get enhanced as we increase the number of barriers, in which case strong level repulsion is verified for as few as $N=3$ particles. 
Our analysis is done for bosons, but can also be extended to systems with a small number of distinguishable fermions. 

\section{Experimental realization}

The experimental realization of our model can be done with a controllable number of interacting atoms trapped in one or several one-dimensional traps. Like experiments based on atoms in optical lattices \cite{2008Blochmany,2012Lewensteinultracold}, such systems allow precision control and find potential applications in quantum engineering and quantum technologies~\cite{2019Sowinskione}.  They are also suitable test beds for addressing the question of the transition from few- to many-body systems~\cite{2010Blumejumping, 2012Blume, 2019Sowinskione}. Our choice of the Kronig-Penney potential is also motivated by recent experiments that have achieved coherent optical lattices with
a sub-wavelength spatial structure that consists of ultra-narrow barriers~\cite{Wang2018}. This technique can create sharp-box potentials and the heights of individual barriers can be varied, which adds a further tunable parameter to the system. 

Our results apply to few-body atomic systems with identical particles that possess spatial wave functions symmetric under particle exchange. Wave functions with this spatial symmetry are clearly relevant for the description of bosons, but our results are equally valid for {\it distinguishable} fermions, i.e.\ identical fermions with internal degrees of freedom like spin. For a system of distinguishable fermions, the antisymmetry required by particle statistics can be carried by the spin or internal wave function. All permitted symmetries for three fermions or bosons (distinguishable or indistinguishable)  are discussed in~\cite{2012Harshman,2014GarciaMarch,2016Harshman}; for more particles in~\cite{harshman_one-dimensional_2016}; for few particles in double or few wells in~\cite{2015GarciaMarch, harshman_infinite_2017}. 

 Due to  losses via three-body re-combination, experiments with a few bosons trapped in the  ultracold regime have an additional difficulty when compared with those with a few fermions. However, for the order of tens of bosons, it was shown in~\cite{2013Bourgain} that one can successfully load dipole traps by means of evaporative cooling. Smaller number of atoms can be loaded in optical lattices~\cite{2010Will}, in  arrays of double wells~\cite{2008Cheinet}, or  in a two-site optical ring, which can be  appropriately reshaped into a Gaussian trap~\cite{2010He}.  An alternative  experimental route leading to cooled atoms trapped in several wells is that of few atoms in optical tweezers, which for a single atom in the ground state was accomplished in~\cite{2012KaufmanPRX}.  In subsequent papers it was experimentally demonstrated the trapping of two $^{87}\mathrm{Rb}$ bosonic atoms in two wells~\cite{ 2014KaufmanScience} or in uniformly filled arrays of traps~\cite{2015LesterPRL}. Recent advances allow the laser cooling of atoms in optical tweezers~\cite{2016Regal}; the trapping of individual atoms in optical tweezer arrays~\cite{2019Saskin}; and even the loading of atoms one by one \cite{2020Schymik,2016Barredo,2016Endres}
in a one-dimensional array~\cite{2016Endres}.

 Another experimental context for the  results in this paper is the ground breaking experiments that showed the accessibility and versatility of  systems with a very small number of interacting fermions.   Reference~\cite{2011SerwaneScience} demonstrated that a deterministic number of ultracold fermions could be extracted from a larger ensemble by applying a tightly confined one-dimensional dimple potential. 
In this experiment, a few $^6\mathrm{Li}$ atoms in the two lowest-energy Zeeman substates were trapped.  The strength of the interactions between atoms in different spin states could be controlled via Feshbach resonance. In a subsequent experiment by the same group, they considered one atom of one species (an impurity) interacting with an increasing number of identical fermions, being able to build a small Fermi sea adding fermions one by one in a controllable manner~\cite{2013Wenzfew}. One can have more than two components in a few fermion system, as in the experiment reported in~\cite{2014Pagano},  where a one dimensional system with a tunable number of spin components was realized.  

In addition to the controllable number of atoms, another ingredient required for the realization of our model is the ability to change the trapping potential, creating double, triple or generally multiwell potentials. The same group that realized few trapped fermions in  \cite{2011SerwaneScience} was also able to trap few fermions in one dimensional double-well \cite{2015Murmanntwo,2019Bergschneiderexperimental} and multi-well systems~\cite{2020BecherPRL}.

\section{Model}

In the simplest case of equally-spaced barriers, the Hamiltonian describing our system takes the form
\begin{subequations}\label{eq:model}
   \begin{equation}
   \frac{1}{\epsilon_1} H(N,W,\tau,\gamma) = T^N  + \tau V^{N,W}  +  \gamma U^N
\end{equation}
where
\begin{eqnarray}
    T^N   &=&  -\sum_{i=1}^N  \frac{\partial^2}{\partial x_i^2}\\
    V^{N,W} &=&  \sum_{i=1}^N \sum_{k=1}^{W-1} \delta(x_i - \pi k/W) \\
    U^N & = & \sum_{\langle ij \rangle} \delta(x_i - x_j). \label{eq:bar} 
    \label{eq:int}
\end{eqnarray}
\end{subequations}
This model realizes a system with $N$ identical interacting particles of mass $m$ trapped in a one-dimensional box of length $L$ that is disrupted by $W-1$ delta-barriers.  
In the equation above, $x_i \in [0,\pi]$ are the positions of the particles scaled by the length $L/\pi$, the energy scale is provided by $\epsilon_1 =\hbar^2\pi^2/(2m L^2)$ (henceforth set to unity), and $\tau \geq 0 $ and $\gamma  \geq 0 $ are unitless parameters describing the barrier strength, and interaction strength, respectively. Here we consider repulsive interactions, as it is the most common scenario for ultracold bosons. 

The goal of our analysis of the Hamiltonian $H(N,W,\tau,\gamma)$ is to understand how the signatures of quantum chaos scale with the number of particles $N$, the number of wells $W$, the barrier strength $\tau$ and the interaction strength $\gamma$. The particularly simple form (\ref{eq:model}) of the Hamiltonian means that it is amenable to analytic and numeric calculations. 

\subsection{Solvable and integrable limiting cases}
\label{sec:solvable}

\begin{figure}
    \centering
        \includegraphics[width=1.1\columnwidth]{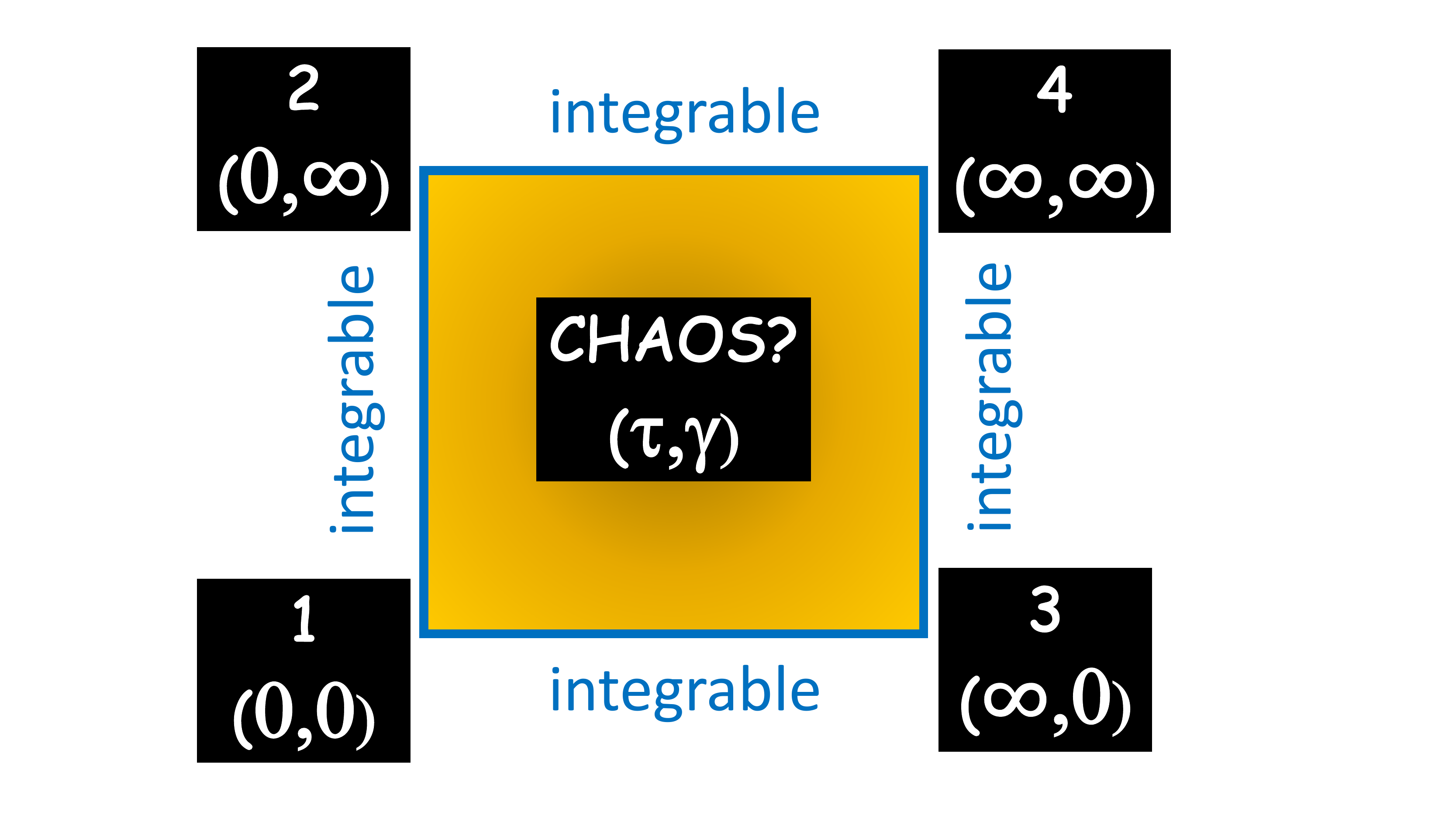}
    \caption{Depiction of the (compactified) $(\tau,\gamma)$ parameter space window $\tau \in [0,\infty)$ and  $\gamma \in [0,\infty)$. At the four corners of the window, the Hamiltonian $H(N,W,\tau,\gamma)$ has exact solutions and is superintegrable. The four edges connecting these corners are integrable models where exact solutions can be found by the solution of coupled transcendental equations. Except for possibly the case of $N=2$, the $H(N,W,\tau,\gamma)$ does not appear integrable or solvable for arbitrary $(\tau,\gamma)$ in this parameter space}
    \label{fig:modelspace}
\end{figure}

For a fixed $N$ and $W$, consider the parameter space window $\tau \in [0,\infty)$ and  $\gamma \in [0,\infty)$, depicted in Fig.~\ref{fig:modelspace}.  An interesting feature of the Hamiltonian $H(N,W,\tau,\gamma)$ is that the model  has exact solutions for four special limiting cases:
\begin{enumerate}
    \item $H(N,W,0,0)=T^N$: $N$ non-interacting identical particles in a one-dimensional infinite square well with width $L$.
    \item  $H(N,W,0,\infty)$: $N$ hard-core identical particles in a one-dimensional infinite square well  with width $L$.
    \item $H(N,W,\infty,0)$: $N$ non-interacting identical particles distributed in $W$ identical one-dimensional infinite square wells with width $L/W$.
    \item $H(N,W,\infty,\infty)$: $N$ hard-core identical particles distributed in $W$ identical one-dimensional infinite square wells  with width $L/W$.
\end{enumerate}
In all of these four cases, the configuration space is sectioned into one or more $N$-dimensional polytopes with high symmetry. Solving for the spectrum is equivalent to solving the Schr\"odinger equation with Dirichlet boundary conditions. Exact solutions for the Schr\"odinger equation in these polytopes can be constructed from symmetrized combinations of one-particle  states using methods of Refs.~\cite{krishnamurthy_exact_1982,turner_quantum_1984,jain_exact_2008,olshanii_exactly_2015}. 

Beyond these four exactly solvable special cases, the Hamiltonian $H(N,W,\tau,\gamma)$ is  also \emph{integrable}  along the four edges of the  $(\tau,\gamma)$ parameter space window (cf.\  Fig.~\ref{fig:modelspace}).  Referring to the four `corner' models denoted above, the integrable limits are:
\begin{itemize}
   \item[$1  \leftrightarrow 2$] $H(N,W,0,\gamma)$:  Without barriers, the Hamiltonian (\ref{eq:model}) for $N$ identical particles in an infinite square well with delta-function interactions is solvable by coordinate Bethe ansatz~\cite{gaudin_boundary_1971, gaudin_bethe_2014, oelkers_bethe_2006}. The energies are the solution of coupled transcendental equations that depend on $N$ and $\gamma$.
   \item[$1 \leftrightarrow 3$] $H(N,W,\tau,0)$: $N$ non-interacting particles in an infinite square well with $W-1$ barriers. The Hamiltonian separates into $N$ identical one-dimensional sub-Hamiltonians. The spectrum for each sub-Hamiltonian is obtained via solution to transcendental equations that depend on $\tau$ \cite{Reshodko_2019}.
    \item[$2  \leftrightarrow 4$] $H(N,W,\tau,\infty)$: In this limit of finite wells with  hard-core interactions, the solutions are Tonks-Girardeau constructions derived from the symmetrized Slater determinants of the solutions of the non-interacting case $H(N,W,\tau,0)$~\cite{1960Girardeau}.
    \item[$3  \leftrightarrow 4$] $H(N,W,\infty,\gamma)$: With infinite barriers, the configuration space fractures into $W^N$ configurations, i.e.~each particle $i\in\{1,\ldots,N\}$ is in a well $j\in\{1,\ldots,W\}$ with width $L/W$. Each of these particle-well configurations is an independent system, since with $\tau \to \infty$ there is no tunneling among wells. Within each configuration, the solutions are similar to the coordinate Bethe ansatz solutions of $H(N,W,0,\gamma)$, except that now there may be different partitions of interacting particles and the size of the well is $L/W$. 
\end{itemize}
 Note that these four integrable models allow us to establish an adiabatic map between the energy levels for the solvable (and superintegrable) models at the corners of parameter space.

\subsection{Symmetries and degeneracies}

To analyze signatures of chaos, the Hilbert space $\mathcal{H}$ must first be decomposed into subspaces with fixed symmetry. The Hamiltonian (\ref{eq:model}) has two symmetries for any interaction strength and barrier strength in the $(\tau,\gamma)$ parameter window. First, it is symmetric under particle permutations $S_N$, i.e.\ any permutation $p\in S_N$ is represented as a linear transformation in configuration space $(x_1,x_2,\ldots,x_N) \to (x_{p_1},x_{p_2},\ldots,x_{p_N})$ that leaves the Hamiltonian (\ref{eq:model}) invariant. The second symmetry is total parity inversion $\Pi$ about the center of the well, implemented as the affine linear transformation  $x_i \to \pi - x_i$ for all $i$. 

These two symmetries allow eigenstates to be classified by irreducible representations of $S_N$  and of parity $\Pi$ and reduce the total Hilbert space $\mathcal{H}$ into subspaces with a given symmetry (cf.~\cite{harshman_one-dimensional_2016}). In this work, we focus on the sector of Hilbert space $\mathcal{H}^{[N]+} \subset \mathcal{H}$ containing states with bosonic symmetry under particle exchange and positive parity.
Note that a special property of delta-function operators $V^{N,W}$ and $U^N$ is that they vanish on certain subspaces of $\mathcal{H}$, because the support of the delta-functions coincides with nodal lines of symmetrized eigenstates of $T^N$. In particular, there is a subspace of $\mathcal{H}^{[N]+}$ upon which $V^{N,W}$ vanishes when $N$ is even  or both $N$ and $W$ are odd. Also, in the limits $\tau \to \infty$ and $\gamma \to \infty$, both $V^{N,W}$ and $U^N$ must vanish on any states with finite energy; in this limit the wave functions must have nodal surfaces that coincide with the support of these operators.

Degeneracies in the spectrum either originate with the symmetries of the Hamiltonian or they are designated `accidental'. 
 Because the only symmetries for generic $(\tau,\gamma)$ are particle exchanges $S_N$ and parity $\Pi$, there should be no degeneracies originating in symmetry in the interior of parameter space. This is because the irreducible representations $[N]+$ for totally symmetric bosonic states are one-dimensional. However, additional symmetries  arise in the limiting cases of zero and infinite strengths $\tau$ or $\gamma$, including the symmetries of separability and system decoupling due to infinite barriers~\cite{harshman_infinite_2017}. These symmetries explain the integrability of the edge models and the superintegrability of the corner models. When $\tau$ and $\gamma$ are both finite, then these additional symmetries are broken and we expect these energy levels that cross at certain parameters along the edges of the model space to repel. Level repulsion, which is a signature of chaos, is thus expected in the interior of this model space (see Fig.~\ref{fig:modelspace}), which is indeed what we verify for $N \geq 3$.

Besides degeneracies originating in symmetries, several other kinds of accidental degeneracies should be possible in our system. The variation of two parameters is sufficient to allow for `diabolic points,' topologically-stable conic degeneracies between two energy levels~\cite{berry_quantizing_1981, berry_diabolical_1984}. These degeneracies can be distinguished from `near misses' by looking at how the wave function transforms when small loops in control space are taken and in subsequent work we will consider such loops. There are  also degeneracies for the solvable corner models arising from the number theory of decomposing integers into sums of squares of integers. Also called Pythagorean degeneracies~\cite{shaw_degeneracy_1974,harshman_infinite_2017}, the density of such degeneracies grows slowly with energy but eventually comes to dominate the spectrum~\cite{berry_quantizing_1981}. Finally, there are other degeneracies characteristic of Bethe-ansatz solvable systems~\cite{heilmann_violation_1971} that should be relevant for  some of the integrable edge models. It is not clear how these additional `accidental' degeneracies at the corners and edges affect the spectrum of the interior of parameter space, but they inform our interpretation of the level statistics presented below  for the $N=2$ case.

\subsection{Density of states}

The density of states of the Hamiltonian $H(N,W,\tau,\gamma)$ is used to interpret some of the numerical results below. Perhaps surprisingly, it is independent of $\tau$ and $\gamma$ to leading order in the energy $E$. The density of states in the sector $\mathcal{H}^{[N]+}$ is

\begin{eqnarray}\label{eq:dos}
    \rho^{[N]+}(E) &=&  \frac{1}{2^{N+2}(N+1)!}\frac{\pi^{N/2}}{\Gamma(N/2+ 1)} E^{N/2-1}\nonumber\\
    && + O\left[N,W,\tau,\gamma\right](E^{N/2 -3/2}),
\end{eqnarray}
where we have used the  notation $O\left[N,W,\tau, \gamma\right]$ to indicate that generally, the coefficient in front of the subleading term proportional to $E^{N/2 -3/2}$ will depend on all the parameters of the Hamiltonian. 

\begin{figure}
    \centering
    \includegraphics*[width=2.7in]{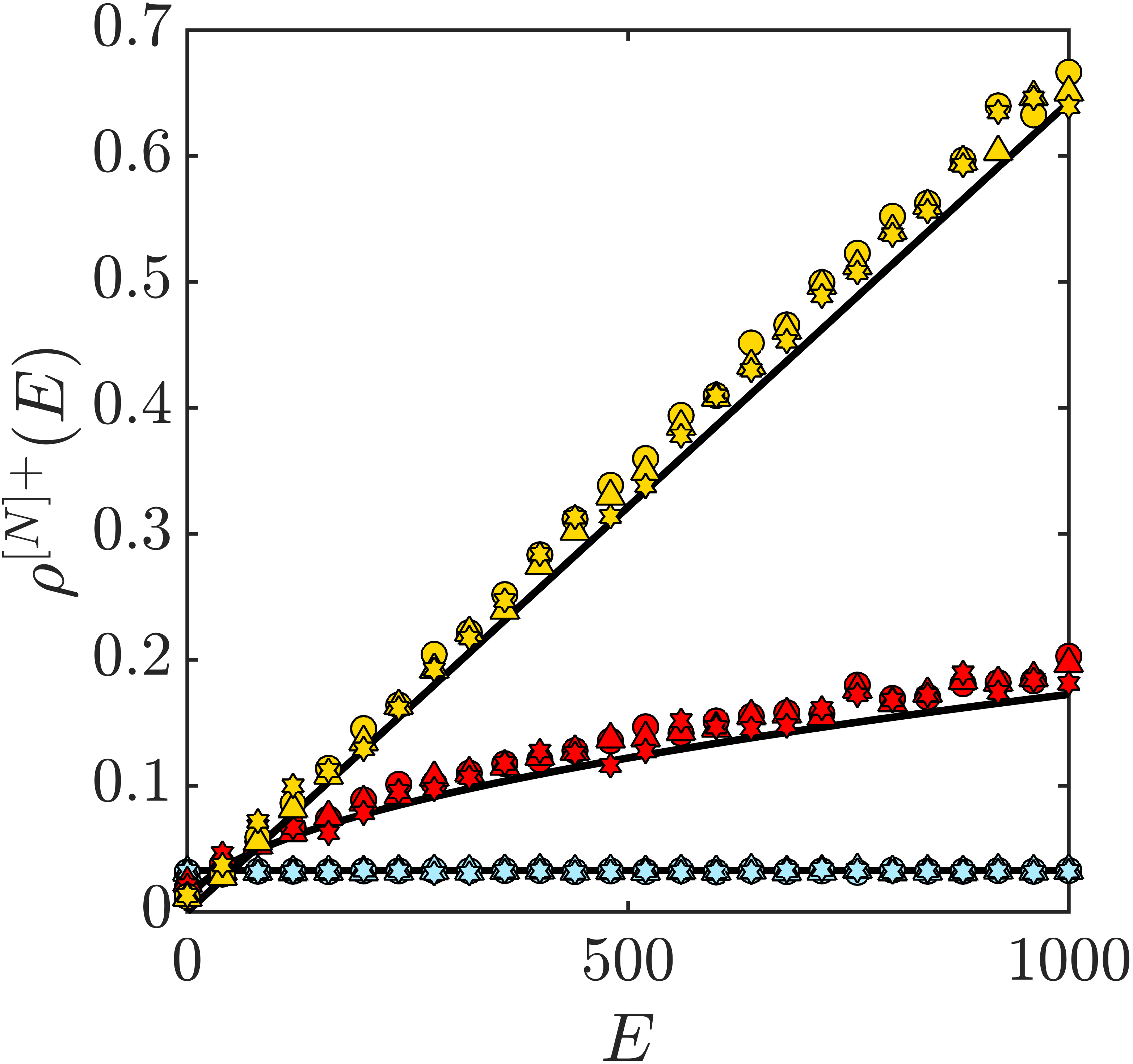} 
    \caption{Density of states for $N=2$ (blue points), $N=3$ (red points) and  $N=4$ (yellow points). Solid lines show the leading term in Eq.~\eqref{eq:dos} for the respective particle numbers. Symbols denote different system parameters: $\gamma=10$ and $\tau=0$ (circles), $\gamma=\tau=10$ and $W=2$ (triangles) and  $\gamma=\tau=10$ and $W=10$ (stars).}
    \label{fig:energy_density}
\end{figure}

We derive Eq.~(\ref{eq:dos}) in the Appendix for the four solvable `corner' models, and infer that it holds for the entire parameter window. More generally, adding one-dimensional delta-functions to an otherwise free problem should not change the density of states. By Weyl's Law~\cite{ivrii_100_2016}, the volume of phase space can be related to the density of states, and boundary conditions (whether Dirichlet, Neumann or hybrid) do not change the phase space volume to leading order. 

In Fig.~\ref{fig:energy_density} we show that the leading term in Eq.~(\ref{eq:dos}) agrees with the scaling of the density of states regardless of number of wells $W$, barrier strength $\tau$ or interaction strength $\gamma$. Because the leading term in Eq.~(\ref{eq:dos}) is the same for the entire parameter window for a given $N$, at least heuristically we expect that the variation in level statistics with the parameters $(\tau,\gamma)$ is governed by the properties of the subleading term (or even lower order terms). In the Appendix, we support this hypothesis by looking at how energy levels change as the model parameters are adiabatically tuned along integrable edge models to connect the spectra of the  solvable corner models.

Note that for $N=2$,  the density of states is a constant and the subleading correction is actually decreasing with energy as $E^{-1/2}$. In Fig.~\ref{fig:energy_density}, all data points for $N=2$ lie on top of the line given by the leading term in Eq.~(\ref{eq:dos}).  For $N=3$, the density of states increases as $E^{1/2}$ and the subleading term is constant. Only for $N=4$ are both the leading and the subleading term of the density of states growing with energy. If the subleading term is important for understanding the density of level crossings as we propose, then this helps to explain why $N=4$ is the threshold when the level statistics conform to the expectations of random matrix theory across such a wide range of parameter space, whereas the $N=3$ only in a limited range.

\section{Indicators of chaos}
\label{sec:indicators}

In the parameter window of interest, finite interactions and finite barrier heights, the Hamiltonian \eqref{eq:model} is diagonalized numerically using a basis consisting of $N$-particle non-interacting eigenstates of $T^N$ in the Hilbert space $\mathcal{H}^{[N]^+}$ \cite{2012Harshman}. In the following we discuss the different indicators of chaos that are used in this work along with the numerical results.

\subsection{Energy spacing distribution}
\label{sec:density}

To study the degree of short-range correlations between the eigenvalues, we use the distribution $P(s)$ of the spacings between neighboring levels obtained after unfolding the spectrum. In generic integrable models, the level spacing distribution is Poissonian, $P_{\rm P}(s) = e^{-s}$, when the energy levels are uncorrelated and not prohibited from crossing, although different shapes emerge for ``picket-fence''-kind of spectra~\cite{Berry1977,Pandey1991,Garcia_March_2018} and systems with an excessive number of degeneracies~\cite{Zangara2013}. For chaotic systems with real and symmetric Hamiltonian matrices, as the one considered here, the level spacing distribution follows the Wigner-Dyson distribution~\cite{MehtaBook,Guhr1998}, $P_{\rm WD}(s) = (\pi s/2) \exp \left( -\pi s^2/4 \right)$, which indicates that the eigenvalues are correlated and repel each other.

\begin{figure}[t!]
    \centering
    \includegraphics[width=\columnwidth]{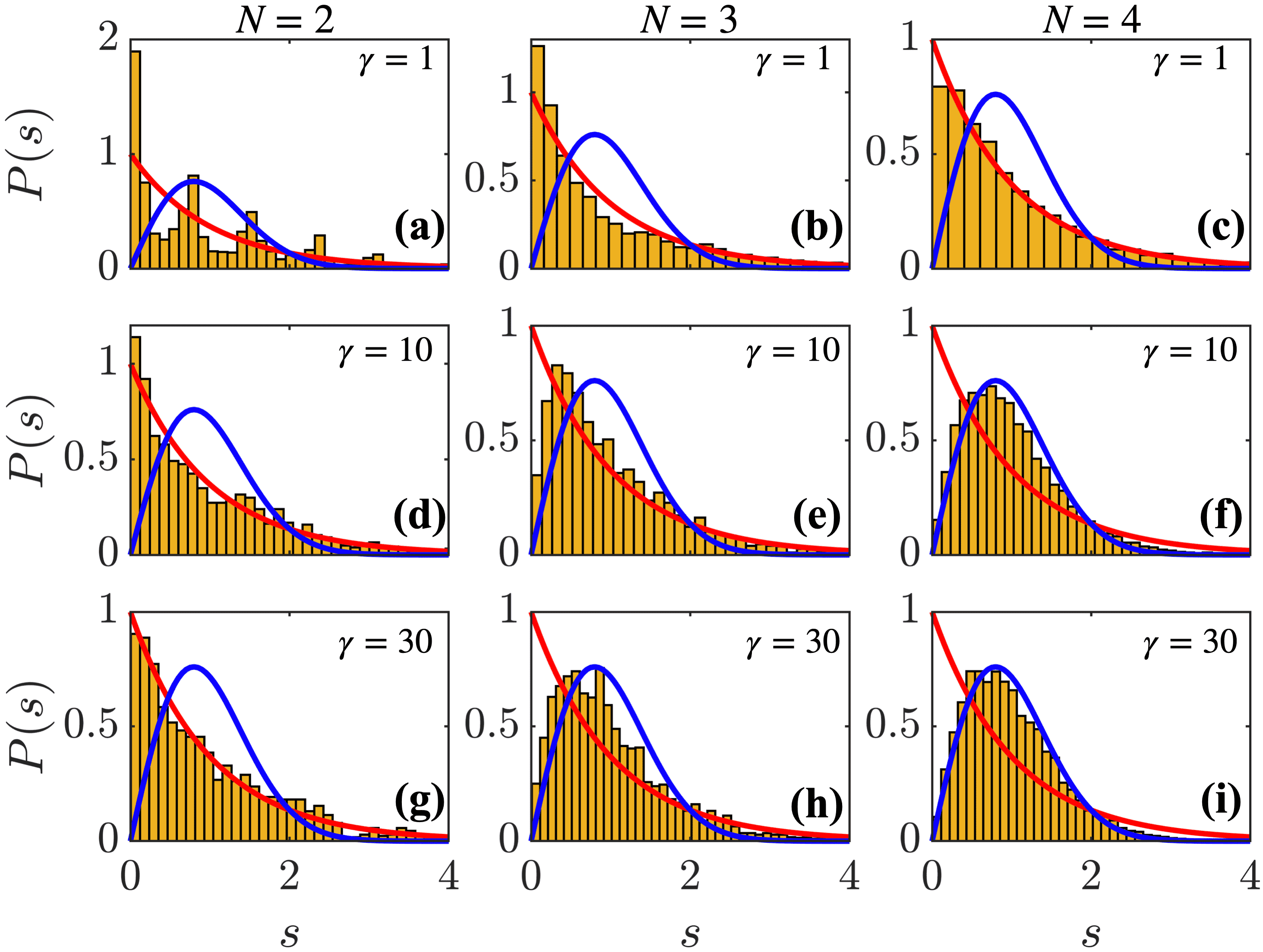}
    \caption{Energy level spacing distributions for (a,d,g) $N=2$ particles, (b,e,h) $N=3$ particles and (c,f,i) $N=4$ particles with different interaction strengths $\gamma$. The barrier height is fixed to $\tau=10$ and the number of wells is $W=2$. The red solid line is the Poissonian distribution and the blue solid line is the Wigner-Dyson distribution.}
    \label{fig:energy_spacing}
\end{figure}

First we investigate the minimal case for chaos in our system, where only one barrier is inserted centrally in the square well, thereby creating a double well potential ($W=2$). 
In Fig.~\ref{fig:energy_spacing} we show the level spacing distributions for $2$, $3$ and $4$ particles for different interactions and at a fixed barrier height of $\tau=10$. 

For weak interactions, $\gamma=1$, a picket fence pattern is noticeable for $N=2$ particles [Fig.~\ref{fig:energy_spacing}~(a)], which indicates non-generic correlations in the energy spectrum~\cite{Berry1977}. Additionally, the peak at $s=0$ signals excessive degeneracies typical for solvable systems~\cite{Zangara2013, heilmann_violation_1971}. For $N=3$, the picket fence structure vanishes and the distribution is closer to Poissonian with some remaining evidence of additional degeneracies at $s=0$ [Fig.~\ref{fig:energy_spacing}~(b)]. For $N=4$, the distribution is also close to Poissonian, except for a slight decrease at $s=0$ that provides evidence for level repulsion already at weak interactions [Fig.~\ref{fig:energy_spacing}~(c)].

At stronger interactions ($\gamma\geq 10$), the distributions for $N=2$ particles in Fig.~\ref{fig:energy_spacing}~(d) and Fig.~\ref{fig:energy_spacing}~(g) become close to Poissonian and level repulsion is not evident. This Poissonian distribution suggests that for the $N=2$ and $W=2$ the Hamiltonian is integrable (or effectively so) for all values of $(\tau,\gamma)$; see the conclusion for a further discussion on this point.
In contrast, some degree of level repulsion is already noticeable for $N=3$ [Fig.~\ref{fig:energy_spacing}~(e) and Fig.~\ref{fig:energy_spacing}~(h)] and the Wigner-Dyson distribution is visible for $N=4$ particles [Fig.~\ref{fig:energy_spacing}~(f) and Fig.~\ref{fig:energy_spacing}~(i)]. 

To explore the crossover of the energy spectrum from Poissonian to Wigner-Dyson and the role of interactions and the potential barrier, we fit our numerical results to the Brody distribution~\cite{Brody1981} (see an alternative in~\cite{Izrailev1990}),
\begin{eqnarray}
P_\beta (s) &=& (\beta+1) b s^{\beta} \exp(-b s^{\beta+1}), \label{Eq:beta} \\ 
b &=&\left[ \Gamma \left( \frac{\beta+2}{\beta+1} \right) \right]^{\beta+1}. \nonumber
\end{eqnarray}
For the Wigner-Dyson distribution, $\beta \sim 1$, while the Poissonian distribution leads to $\beta \sim 0$.

\begin{figure}[t]
    \centering
    \includegraphics[width=\columnwidth]{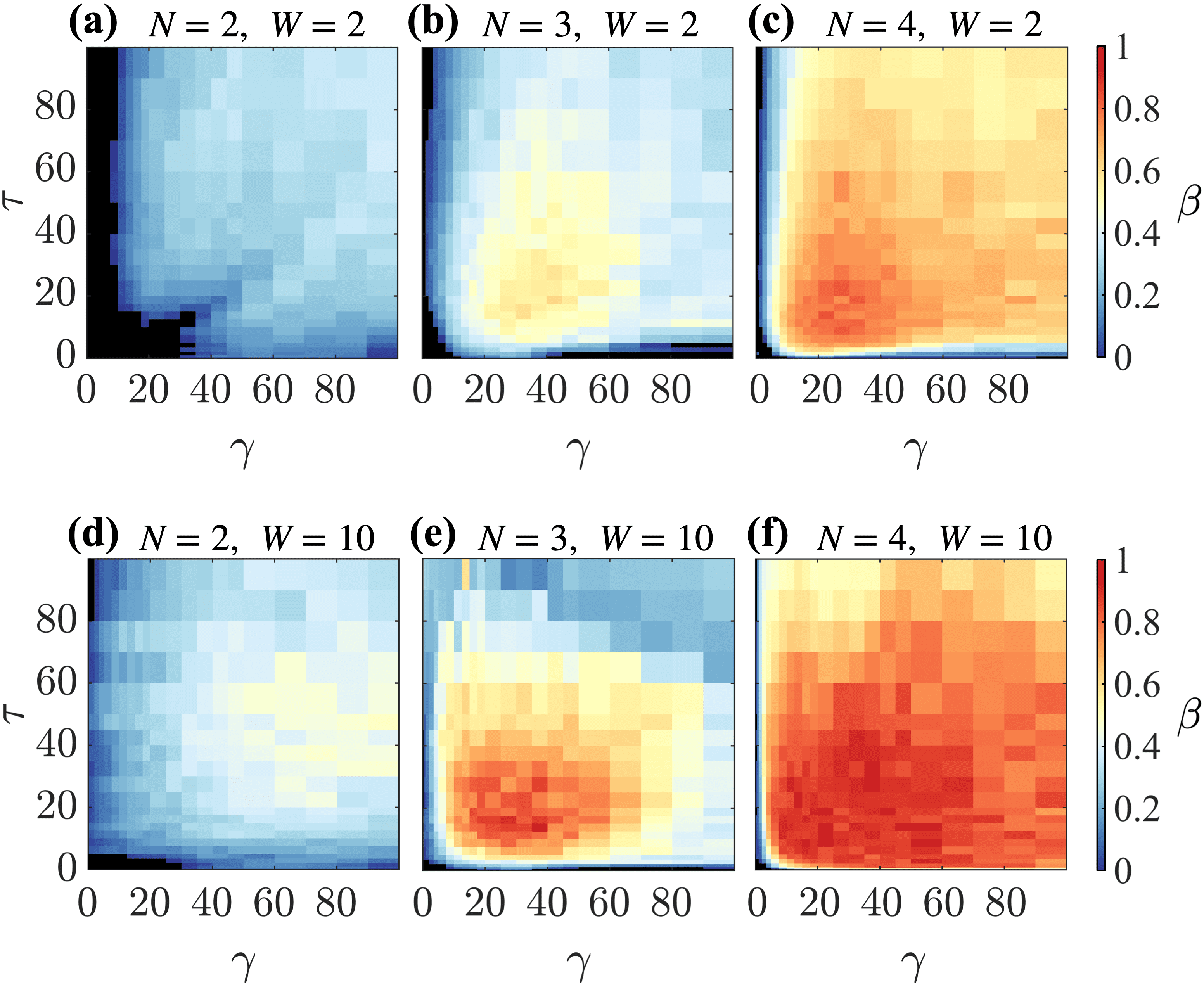}
    \caption{Brody distribution parameter $\beta$ as a function of $\tau$ and $\gamma$ for (a,d) $N=2$, (b,e) $N=3$, and (c,f) $N=4$ particles. The number of wells are (a-c) $W=2$ and (d-f) $W=10$. For $W=2$ wells the maximum $\beta_{max}=0.8269$ is found for $N=4$ particles at $\gamma=20$ and $\tau=12.5$, while for $W=10$, we find $\beta_{max}=0.9512$ for $N=4$ particles at $\gamma=45$ and $\tau=12.5$. Black areas show $\beta<0$ indicating degeneracies and picket-fence spectra.}
    \label{fig:beta_W2_W10}
\end{figure}

Close to the integrability limits, vanishing interactions ($\gamma \approx 0$) with finite barriers ($\tau\neq 0$) or finite interactions ($\gamma \neq 0$) with vanishing barrier height ($\tau \approx 0$), the energy spacing distributions can display a picket-fence or quasi-Poissonian distribution as discussed in Fig.~\ref{fig:energy_spacing} for $N=2$, which results in $\beta<0$. We highlight these non-generic regions as black in Figs.~\ref{fig:beta_W2_W10}~(a)-(f), showing that they are more prevalent at smaller particle numbers, while their footprint in the parameter space almost vanishes for $N=4$. As we increase the interactions ($\gamma \gg 0$) and the barrier heights ($\tau \gg 0$), the degeneracies of  the $N=2$ system are destroyed and the level spacing distributions become closer to Poissonian for a large region of the parameter space with $\beta$ remaining below $0.35$.

For more particles, the distributions are different, with indications of energy level repulsion emerging for $N=3$ in the region $15 \lesssim \gamma \lesssim 45$ and $5 \lesssim \tau \lesssim 35$ where $\beta\sim 0.5$ [Fig.~\ref{fig:beta_W2_W10}~(b)]. For $N=4$  the fitting parameter $\beta$ in almost the entire parameter space is larger than $0.5$, with the areas of integrability confined to the edges of the $(\tau,\gamma)$ parameter space [Fig.~\ref{fig:beta_W2_W10}~(c)]. In fact, we find the maximum to be around $\beta \sim 0.83$ at $\gamma=20$ and $\tau=12.5$, indicating that the energy spacing distribution gets close to Wigner-Dyson. 

In Figs.~\ref{fig:beta_W2_W10}~(d)-(f) we also consider multiple wells, with $W=10$. For more wells, the density of states does not change to leading order in $\tau$, but a greater proportion of the eigenstates of $T^N$ feel the effect of the barriers more acutely. For example, for $N=2$ and $W=2$, the barrier potential $V^{N,W}$ vanishes on half of the eigenstates of $T^N$ in $\mathcal{H}^{[N]+}$, whereas for $N=2$ and $W=10$, the barrier potential  $V^{N,W}$ only  vanishes on one tenth of the eigenstates of $T^N$ in $\mathcal{H}^{[N]+}$.

Comparing the case of $W=10$ to $W=2$ in Fig.~\ref{fig:beta_W2_W10}, we see that larger values of $\beta$ are indeed found for $W=10$ and the region of the parameter space where $\beta$ is large has increased significantly for both $N=3$ and $N=4$. This suggests that in systems with a large number of wells, the energy level repulsion is enhanced and chaos could be observed for systems with as few as $N=3$ particles. However, for $N=2$, we find a maximum of $\beta\approx0.5$ and this value does not approach the chaotic limit by increasing the number of wells. As before, the case of only two particles resists the transition to chaos. In Sect.~\ref{sec:barriers} below, we discuss in more detail how the signatures of chaos change as the number of barriers is increased.

\begin{figure}[t]
    \centering
    \includegraphics[width=\columnwidth]{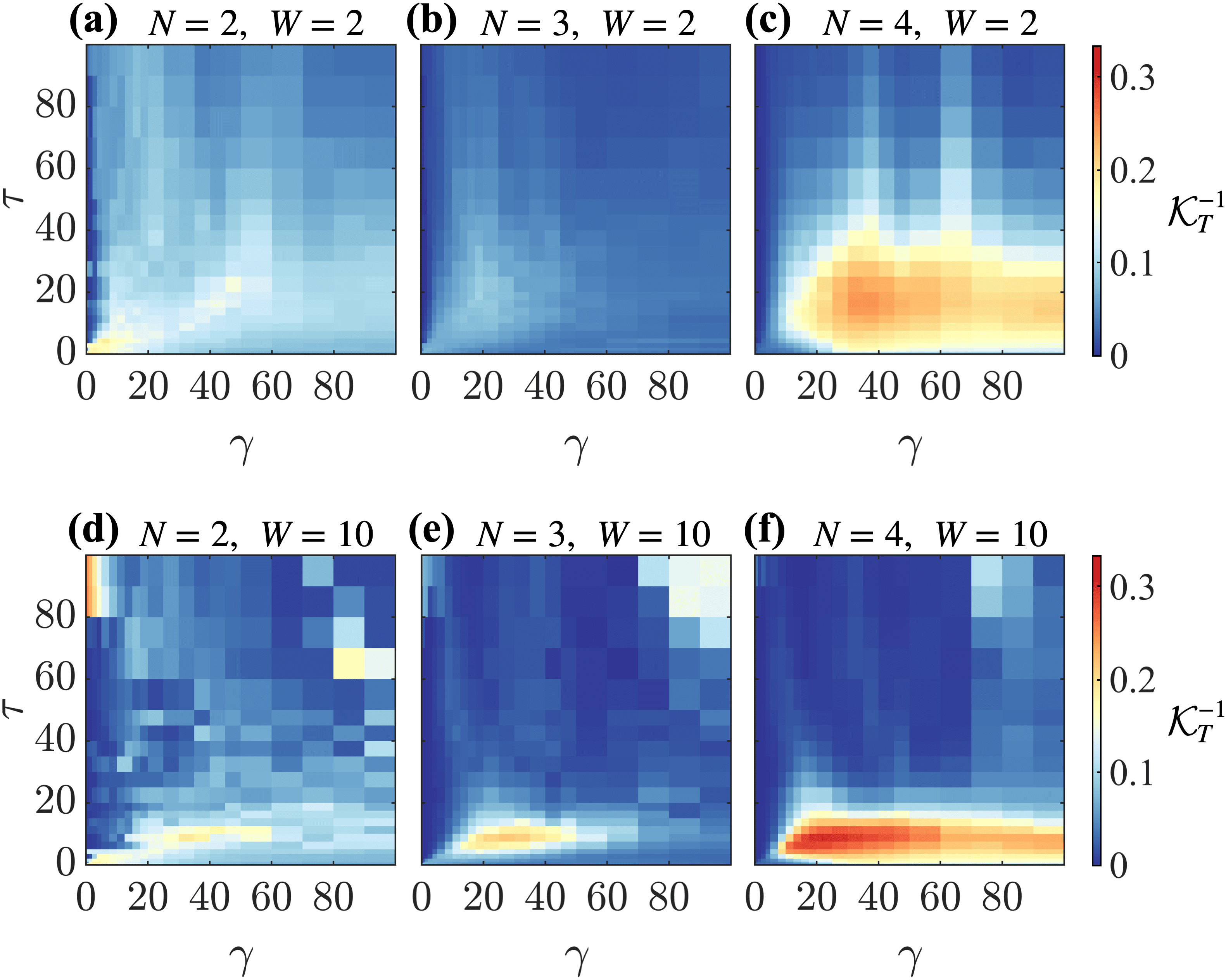}
    \caption{Inverse of the kurtosis $\mathcal{K}^{-1}_{T}$ as a function of $\tau$ and $\gamma$ for (a,d) $N=2$ (b,e) $N=3$ and (c,f) $N=4$ particles. The number of wells are (a-c) $W=2$ and (d-f) $W=10$. For $W=2$ the minimum kurtosis is $4.3592$ at $\gamma=30$ and $\tau=15$, while for $W=10$ the minimum kurtosis is $3.3274$ at $\gamma=20$ and $\tau=7.5$.}
    \label{fig:kurt_W2_W10}
\end{figure}

\begin{figure*}
    \centering
    \includegraphics[width=2\columnwidth]{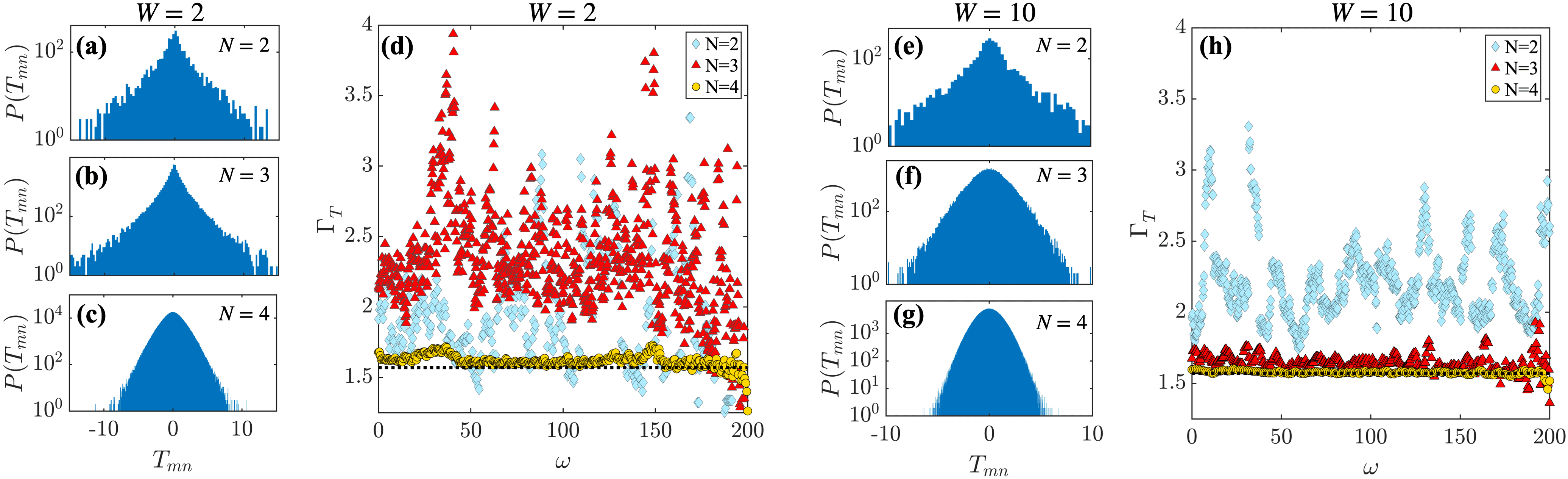}
    \caption{Off-diagonal elements of the kinetic energy operator $T_{mn}$ for (a-d) $W=2$ wells and (e-h) $W=10$ wells. The probability distributions in the respective potentials are shown for (a,e) $N=2$ (b,f) $N=3$ and (c,g) $N=4$. (d,h) The ratio $\Gamma_{T}$ as a function of $\omega=|E_m-E_n|$ in an energy window at $E_{\rm{mid}}=700$ and width $\Delta E=100$. The interaction strengths and barrier heights are chosen at the point of minimum kurtosis for $4$ particles in Fig.\ref{fig:kurt_W2_W10}(c,f), namely $\gamma=30$ and $\tau=15$ for $W=2$, and $\gamma=20$ and $\tau=7.5$ for $W=10$. }
    \label{fig:off_diags}
\end{figure*}

\subsection{Off-diagonal ETH}
Quantum chaos ensures the validity of the ETH, so we can also use the indicators of ETH to detect the transition to chaos. Two conditions need to be satisfied for a few-body observable $O$, evolving according to
\begin{eqnarray}
O(t) &=& \sum_{m \neq n}  C_{m}^* C_{n}   e^{- i (E_{n} - E_{m})t}  O_{m n}\nonumber\\
&&{} + \sum_{m}  |C_{m}|^2  O_{m m} ,
\label{Eq:O}
\end{eqnarray}
to reach thermal equilibrium. In the equation above, $C_{m} =\langle m |\Psi(0) \rangle$ is the overlap between the eigenstate $|m \rangle$ of the Hamiltonian that describes the system and the initial state $|\Psi(0) \rangle$, and $O_{m n} = \langle m |\hat{O} |n \rangle$. The first condition is that of equilibration, which depends on the first term on the right-hand side of Eq.~(\ref{Eq:O}). At long times, due to the lack of degeneracies of chaotic Hamiltonians, the small values of the coefficients $C_{m}$'s obtained when the systems is quenched far from equilibrium, and the small values of the off-diagonal elements $O_{m n}$'s caused by the chaotic eigenstates, the first term in Eq.~(\ref{Eq:O}) leads to small fluctuations that decrease with system size and cancel out on average. So apart from small fluctuations, the observable reaches its infinite-time average $\sum_{m}  |C_{m}|^2  O_{m m} $. The second condition is that this infinite-time average approaches the thermodynamic average as the system size increases, confirming that the equilibrium is indeed thermal. These two steps are usually referred to as off-diagonal- and diagonal-ETH, respectively.

Here, we consider the off-diagonal-ETH to detect the transition to chaos. The distribution of $O_{m n}$ in chaotic (thermalizing) systems is Gaussian~\cite{Beugeling2015,LeBlond2019,Brenes2020,Brenes2020b,Santos2020,Brenes2021}, reflecting the chaotic structure of the eigenstates, while other forms emerge for integrable models~\cite{LeBlond2019}. To quantify how close the distribution of $O_{m n}$ is to a Gaussian, we use the kurtosis,
\begin{equation}
\mathcal{K}_{\hat{O}}= \frac{  \langle( O_{m n} -\langle  O_{m n} \rangle)^{4}\rangle }{\sigma^{4}},
\label{eq:kurtosis}
\end{equation}
where $\langle . \rangle$ indicates the average over all pairs of eigenstates $|m\rangle \neq |n\rangle$ and $\sigma$ is the standard deviation of the distribution. For Gaussian distributions the kurtosis is given by $\mathcal{K}_{\hat{O}}=3$. 

In our calculations of $\mathcal{K}_{\hat{O}}$, we consider the kinetic energy operator $T^N$ (in the following we drop the superscript to simplify the notation). To ensure no effects from the bottom edge of the spectrum, we choose a energy window far from the ground state and include only states within $E_m \in [E_{\rm{mid}}-\Delta E, E_{\rm{mid}}+\Delta E]$, where the center of the energy window is high in the spectrum at $E_{\rm{mid}}=700$ and its width is $\Delta E=100$. In Figs.~\ref{fig:kurt_W2_W10}~(a)-(f) we show the inverse of the kurtosis, $1/\mathcal{K}_{T}$, with the maximal value $1/3$ indicating a Gaussian distribution and therefore the presence of chaos. 

For $N=4$ particles and $W=2$ [Fig.~\ref{fig:kurt_W2_W10}~(c)], the distribution is close to Gaussian for barrier heights $1 \lesssim \tau\lesssim 30$ and interactions $20\lesssim \gamma \lesssim 60$, which coincides with the region of $\beta>0.8$ in Fig.~\ref{fig:beta_W2_W10}~(c). We choose the minimum value of the kurtosis in Fig.~\ref{fig:kurt_W2_W10}~(c) and show the Gaussian probability distribution of  $T_{mn}$ in Fig.~\ref{fig:off_diags}~(c). In contrast, for lower particle numbers, $N=2,3$, the distribution of the off-diagonal elements $T_{mn}$ is sharply peaked, as seen in Fig.~\ref{fig:off_diags}~(a) and Fig.~\ref{fig:off_diags}~(b), which is consistent with the kurtosis having values $\mathcal{K}_{T}\gg 3$ in Fig.~\ref{fig:kurt_W2_W10}~(a) and Fig.~\ref{fig:kurt_W2_W10}~(b).

The kurtosis shows a similar enhancement due to the presence of more barriers, attaining a minimum of $\mathcal{K}_{T}\approx 3.3$ for $N=4$ particles in $W=10$ wells [Fig.~\ref{fig:kurt_W2_W10}~(f)]. Indeed, for $1 \lesssim\tau<20$ and a broad range of interactions, the kurtosis is close to 3. Increasing the number of barriers also moves the band of minimal kurtosis to lower values of $\tau$, as lower barrier heights are necessary to retain the competition with the inter-particle interactions. In a similar region of the parameter space, there is also a visible minimum of the kurtosis for $N=3$ particles [Fig.~\ref{fig:kurt_W2_W10}~(e)]. In fact, when taking the interaction and barrier height which give the minimum value of kurtosis [$\gamma=10$ and $\tau=7.5$ for $N=4$], there are distinct Gaussian probability distributions for both $N=4$ and $N=3$ particles [see Figs.~\ref{fig:off_diags}~(f,g)]. However, for $N=2$ the off-diagonal elements of the kinetic energy  operator $T_{mn}$ do not indicate the presence of chaos in either the kurtosis [Fig.~\ref{fig:kurt_W2_W10}~(d)] or the probability distribution [Fig.~\ref{fig:off_diags}~(e)].

To quantify how the off-diagonal elements $T_{mn}$ behave as a function of the energy difference $\omega=|E_m-E_n|$ we also show in Fig.~\ref{fig:off_diags} the ratio
\begin{equation}
    \Gamma_{T}=\frac{\overline{\vert T_{mn} \vert^2}}{\overline{\vert T_{mn} \vert}^2} ,
\end{equation}
which is equal to $\pi/2$ for a Gaussian distribution. In Fig.~\ref{fig:off_diags}~(d) and Fig.~\ref{fig:off_diags}~(h) this ratio is shown for $W=2$ and $W=10$ wells, respectively, for the same Hamiltonian parameters that gave the lowest values of the kurtosis discussed in the preceding paragraph. For $W=2$ wells, the values of $\Gamma_{T}$ for $N=4$ sit very close to $\pi/2$ over a large range of $\omega$, confirming that the Gaussianity of the distribution is preserved at different energy spacings. The ratio for $N=2,3$ has a large variance with the majority of the points being far from $\pi/2$, which is indicative of the peaked distributions in Fig.~\ref{fig:off_diags}~(a) and Fig.~\ref{fig:off_diags}~(b). For $W=10$ wells, we find that not only $\Gamma_{T}\approx \pi/2$ for $N=4$ particles, but also the $N=3$ system approaches this result. This suggests that the $N=3$ system can be tuned between integrability and chaos by changing the trapping potential.

\subsection{Survival Probability}

Spectral correlations get manifested also in the evolution of the survival probability,
\begin{eqnarray}
\langle S_P(t)  \rangle &=&  \langle \left|\left<\Psi (0)|\Psi (t)\right>\right|^2  \rangle \nonumber\\
&=& \langle \sum_{n,m} |C_n|^2 |C_m|^2 e^{- i(E_n-E_m)t}\rangle,
\label{eq:SP}
\end{eqnarray}
in the form of what is known as the correlation hole~\cite{Leviandier1986,Pique1987,Guhr1990,Wilkie1991,Hartmann1991,Lombardi1993,Alhassid1992,Leyvraz2013,Torres2017,Torres2017Philo,Torres2018,Schiulaz2019,Lerma2019,Cruz2020}, recently referred to also as the ``ramp''~\cite{Cotler2017GUE}. The correlation hole corresponds to a dip below $\langle S_{\infty} \rangle = \langle \sum_{n} |C_n|^4 \rangle$, which is the infinite-time average (saturation value)  of $\langle S_P(t)  \rangle$. This dip emerges also in the spectral form factor $\langle \sum_{n,m}  e^{- i(E_n-E_m)t}\rangle$, but contrary to this one,  the survival probability is a true dynamical quantity. In the equation above, $\left< . \right>$ indicates averages. The survival probability is non-self-averaging~\cite{Prange1997,Schiulaz2020}, so the correlation hole is not visible unless averages are performed. They can be done over initial states, disorder realizations, or, as in our case, they correspond to moving time averages. The correlation hole detects short- and long-range correlations in the spectrum, and in addition, it does not require unfolding the spectrum or separating it by symmetries~\cite{Cruz2020,Santos2020}. In cold atom systems the survival probability is commonly used to probe the non-equilibrium dynamics of few- \cite{Campbell2014,Keller2016} and many-body systems \cite{Goold2011,Knap2012,Chenu2019,Fogarty2020}, and can be experimentally measured using interferometric techniques \cite{Cetina96}.

\begin{figure*}
    \centering
    \includegraphics[width=1.9\columnwidth]{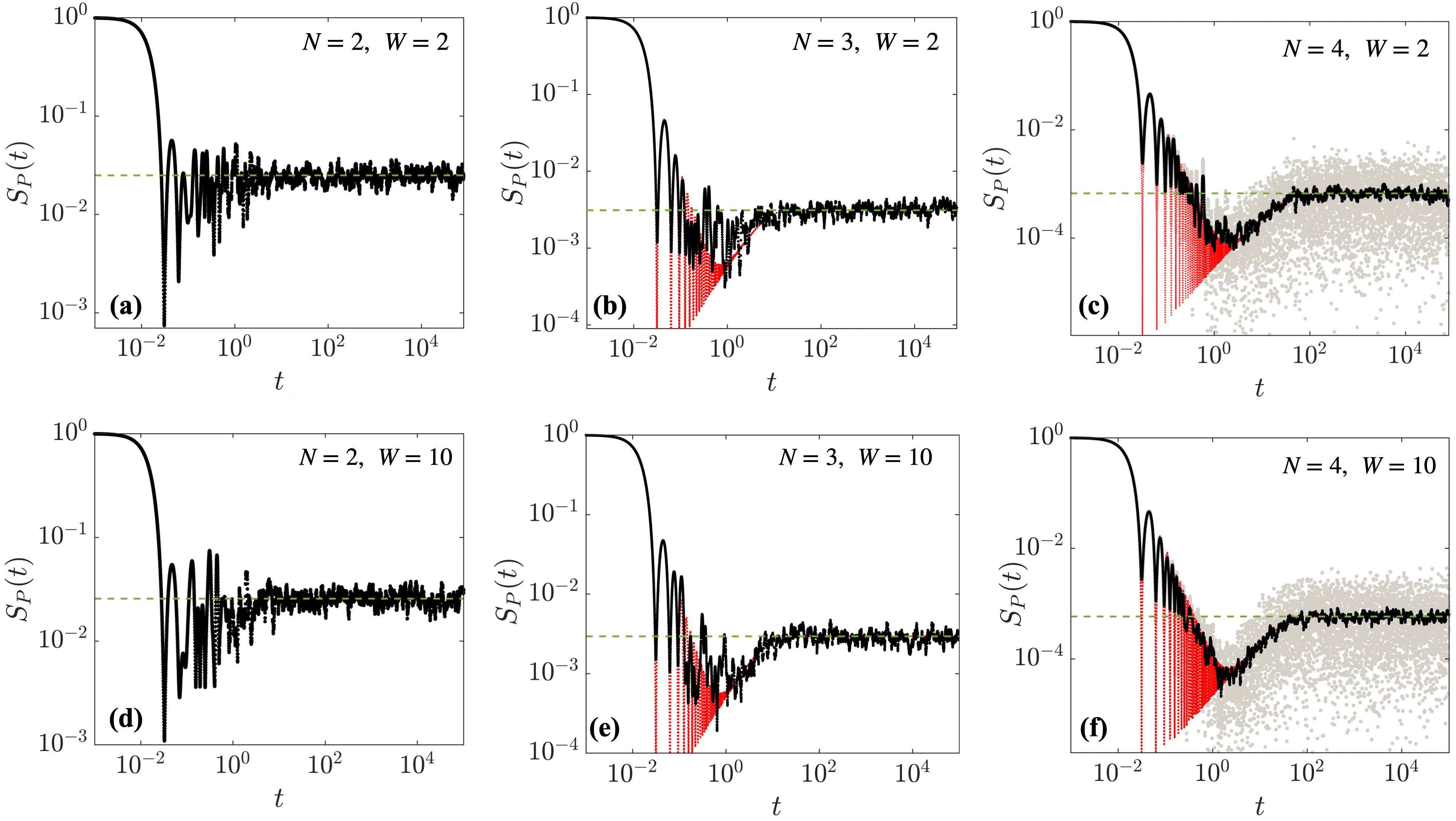}
    \caption{Survival probability of (a,d) $2$ particles (b,e), $3$ particles, and (c,f) $4$ particles. The number of wells is (a-c) $W=2$ and (d-f) $W=10$. The initial state is a square distribution centered at $E_{\rm{mid}}=700$ and of width $\Delta E=100$. Black lines are the moving time averages on the logarithmic timescale 
    with window $[\log_{10}t-\Delta t, \log_{10}t+\Delta t]$ and $\Delta t=0.02$.  Green horizontal dashed line indicates the infinite time average of the survival probability $S_{\infty}$. For $N=3$ and $4$, red lines represent the analytic solution. For $N=4$, the grey points represent the unaveraged value for comparison. The parameters used for each row correspond to those that give the minimum kurtosis for $4$ particles in Fig.\ref{fig:kurt_W2_W10}, namely $\gamma=30$ and $\tau=15$ for $W=2$, and $\gamma=20$ and $\tau=7.5$ for $W=10$.}
    \label{fig:M2_ODs}
\end{figure*}

We take an initial state that has a homogeneous probability distribution centered at $E_{\rm{mid}}$ inside an energy window of width $\Delta E$
\[ 
\rho(E)=\left\{ \!\!\!\! \begin{array}{cl} 
\frac{1}{2\Delta E} & \mbox{for}\ E \in [E_{\rm{mid}}-\Delta E, E_{\rm{mid}}+\Delta E] \\
0 & \mbox{otherwise.}
\end{array}
\right.
\]
In \cite{Torres2018,Schiulaz2019,Lerma2019} a general analytic solution was derived for the survival probability for chaotic systems. For the square distribution used here, this solution is given by
\begin{equation}
    S_P(t)  = \frac{1-S_{\infty}}{\eta-1}\left[ \eta \frac{\sin^2(\Delta E \, t)}{(\Delta E\, t)^2} - b_2 \left( \frac{\Delta E \, t}{\pi \eta} \right) \right]+S_{\infty}\;.
    \label{eq:SprobA}
\end{equation}
Here,  $\eta$ is the number of energy eigenvectors in the energy window $\Delta E$. The initial decay of the survival probability is captured by the first term in the brackets in Eq.~\eqref{eq:SprobA}, after which the dynamics is described by the two-level form factor
\begin{eqnarray*}
b_2(\bar{t}) &=&\left[1-2\bar{t} + \bar{t} \ln (2\bar{t}+1) \right] \Theta(1-\bar{t}) \\
&&{}+\left[ \bar{t}\ln \left(\frac{2 \bar{t}+1}{2\bar{t}-1}\right)-1  \right] \Theta(\bar{t}-1),
\end{eqnarray*}
with $\Theta(\bar{t})$ the Heaviside step function. 

In Fig.~\ref{fig:M2_ODs} the survival probability is shown for different number of particles and number of wells (grey points). The moving time average (black lines) smooths the data and allows us to identify the correlation hole with its ramp towards $S_{\infty}$. For $W=2$ and $N=2,3$ [Figs.~\ref{fig:M2_ODs}~(a,b)] there is no obvious indication of a dip below $S_{\infty}$ that could be described by the two-level form factor. 
For $N=4$ [Fig.~\ref{fig:M2_ODs}~(c)], a noticeable dip manifests below the saturation point and follows closely the analytic solution in Eq.~(\ref{eq:SprobA}) (red lines). This is indicative of chaotic behavior. In the $W=10$ well system the correlation hole for $N=4$ is even more pronounced [Fig.~\ref{fig:M2_ODs}~(f)] with the initial decay and subsequent ramp of $S_P(t)$ matching precisely the analytics. For $N=3$ and $W=10$ [Fig.~\ref{fig:M2_ODs}~(e)], partial revivals still obscure the minimum of the correlation hole, but the ramp towards saturation can be seen to follow the analytic results.

\subsection{Dependence on the number of wells}
\label{sec:barriers}

\begin{figure}
    \centering
    \includegraphics[width=\columnwidth]{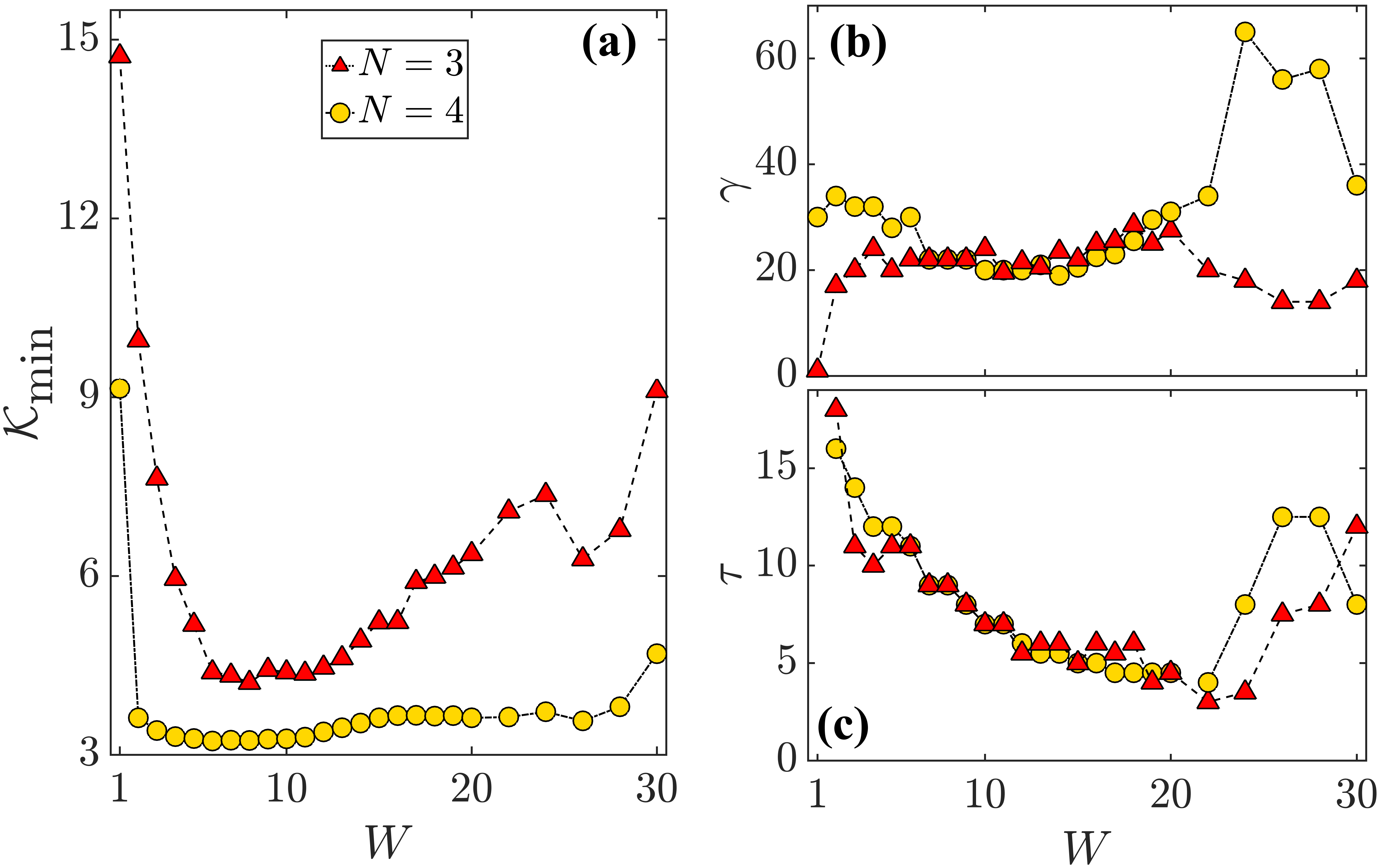}
    \caption{(a) Minimum of the kurtosis as a function of the number of wells for $N=3$ (red triangles) and $N=4$ (yellow circles). Kurtosis is calculated in an energy window of width $\Delta E=100$ around $E_{\rm{mid}}=700$. (b) Interaction and (c) barrier height corresponding to the points of minimum kurtosis in (a).}
    \label{fig:Beta_Kurt_fnM}
\end{figure}

For a more systematic analysis of the onset of chaos as the number of wells in the system is increased, we show in Fig.~\ref{fig:Beta_Kurt_fnM}~(a) the minimum of the kurtosis $\mathcal{K}_{\rm{min}}=\text{min}[\mathcal{K}_{T}]$ in the range $\gamma,\tau \in [0, 100]$. Here we fix the size of the box $L$ and focus on $N=3$ and $N=4$ particles. For both cases the number of wells dictates the emergence of chaos, but to different degrees. For the trivial case of $W=1$ (no barriers) the kurtosis is large and both $N=3$ and $N=4$ are integrable. For $W>1$ the kurtosis of $N=4$ takes low values, $\mathcal{K}_{\rm{min}}\sim 3$, indicating chaos, essentially irrespective of the number of wells when $W \lesssim 13$. However, for $N=3$ the inclusion of more barriers causes a more subtle change to the kurtosis, which decreases slowly as more barriers are introduced, attaining a minimum of $\mathcal{K}_{\rm{min}}\sim 4$ in the region of $6\lesssim W \lesssim 13$. It is in this region that the $N=3$ system displays chaotic signatures as discussed in the previous sections. Interestingly, increasing the number of wells further ($W>13$) results in an increase of the kurtosis and indications of chaos are diminished. A similar tendency is seen for $N=4$, albeit in a less drastic manner, as the kurtosis increases at a lower rate. For both $N=3$ and $N=4$ we expect that as the limit of $N/W\rightarrow 0$ is approached the contest between the ordering of the particles in the wells and their interactions is reduced and that the system slowly returns to what we would see in the continuum: particles in a box \cite{Rigol2005}.

In Fig.~\ref{fig:Beta_Kurt_fnM}~(b) and Fig.~\ref{fig:Beta_Kurt_fnM}~(c) we show the optimal interactions and  barrier heights for achieving the minimum kurtosis shown in Fig.~\ref{fig:Beta_Kurt_fnM}~(a). Figure~\ref{fig:Beta_Kurt_fnM}~(c) shows that the optimal barrier strengths $\tau$ for $N=3$ and $N=4$ have close agreement for all $W$, and that these values decrease with increasing number of wells (for $W \leq 20$). This reduction in $\tau$ is necessary to preserve the competition between barriers and interactions, as when the number of barriers is increased the impact of $\tau$ is magnified. This effect can be seen in the shift of the chaotic region in Fig.~\ref{fig:kurt_W2_W10}~(c) and (f). Fig.~\ref{fig:Beta_Kurt_fnM}~(b) shows that the optimal interaction strengths $\gamma$ found for both $N=3$ and $N=4$ converge to a similar value in the region $7 \lesssim W \lesssim 20$, which encompasses the areas of low kurtosis in Fig.~\ref{fig:Beta_Kurt_fnM}~(a). This suggests that both $N=3$ and $N=4$ particles are chaotic in the same regions of the parameter space $[\tau,\gamma]$. Increasing the number of wells beyond $W=20$ breaks this trend and the parameters for $N=3$ and $N=4$ diverge as the indications of chaos are lost.

\section{Discussion and Conclusion}

To summarize, between $N=2$ and $N=4$ our model makes a clear transition. For $N=4$ particles and just $W = 2$ wells, there are clear  signatures of chaos when both the barrier and the interactions have finite strength. The density of states grows rapidly with energy, and numerical analysis gives evidence of the highly-correlated spectrum typical of random matrices, of the validity of the off-diagonal ETH, and of a clear ramp in the survival probability. As the number of wells is increased, evidence for the onset of chaos becomes more robust, even for weaker barriers and interactions.

In contrast, for $N=2$, the spectrum is numerically indistinguishable from an integrable system throughout the parameter window for $W=2$, and deviates only slightly from this as the number of wells is increased. 
Several possible origins for the non-chaotic behavior of $N=2$ can be hypothesized, including some undiscovered integral of motion, some sort of partial integrability of a subset of eigenstates of the Hamiltonian, a proliferation of accidental degeneracies from several sources that create the appearance of integrability, or a combination of all these.  These possibilities will be considered in more detail in a subsequent work. For now, we note with some irony that integrability is much less generic and more complicated than chaos.

The intermediate case of $N=3$ stands at the ragged edge where symmetry and integrability dissolve into chaos and randomness. By tuning the interaction and barrier parameters and increasing the number of barriers, the full gamut of possibilities can be realized on the same small atomic system. The $N=3$ case has some common aspects with the $N=2$ case and other features similar to $N=4$. For example, the density of states grows with the energy like the $N=4$ and unlike the $N=2$ case, although it grows sublinearly while the $N=4$ case grows linearly. 
On the other hand, the probability distributions for the off-diagonal elements of the kinetic energy are closer to the $N=2$ than to the $N=4$ case for only two wells, but when increasing the number of wells, it gets closer to the Gaussian distribution characteristic of the $N=4$ case.  
The variety of possible scenarios for $N=3$ atoms is most clear in Fig.~\ref{fig:Beta_Kurt_fnM}~(a), where the 
degree of chaoticity measured with the minimum kurtosis 
$\mathcal{K_\mathrm{min}}$ makes sweeping changes as the number of wells is increased. 

As we discuss above, current experiments with subwavelength lattice potentials and deterministically prepared few-body states provide the ideal platform to probe this boundary between integrability and chaos. Furthermore, these small atomic systems have been proposed as the working units of larger quantum information processing devices and protocols. Therefore, understanding their information, control, and entanglement properties in these different regimes becomes important.

For example, this model lends itself naturally to ``digitization''. The number of particles in each well becomes a useful observable in the infinite barrier limit; they are integrals of motion in fact. Similarly, coherent superpositions of eigenstates of these or other integrals of motion in the limiting edge models could be used for storing quantum information (c.f.~\cite{olshanii_creating_2018}).
A quench from an integrable limit to the chaotic parameter regime would break these integrals of motion and effectively scramble the information held in the initial state. 

An important extension of this work that is experimentally relevant 
is determining how sensitive our results are to the idealizations of delta-barriers, precisely symmetric positions and uniform barrier height. We expect that small deviations of the periodic Kronig-Penney lattice, such as finite width barriers, would not significantly alter the chaotic regions of our system. The versatility of our model also allows us to explore the possibility of emergent integrability~\cite{Abanin2019} when more disorder is introduced into the system via non-regularly spaced barriers or barriers of different heights. Both aspects are inspiring and we leave them for future research. Another interesting scenario which we have not explored is the case in which the interactions are attractive, whereby the system is now furnished with a bound state and whose limit at infinite interactions is the so-called super Tonks-Girardeau gas~\cite{Astra2005,Batchelor_2005,Haller1224}.


\begin{acknowledgments}
The authors thank M. Olshanii, T. Busch, A. Fabra and M. Boubakour for insights on integrability and conversations about chaos.  TF acknowledges support from JSPS KAKENHI-21K13856 and the Okinawa Institute of Science and Technology Graduate University. We are grateful for the help and support provided by the Scientific Computing and Data Analysis section of Research Support Division at OIST. LFS was supported by the NSF grant No. DMR-1936006. M.A.G.M.  acknowledges  funding  from  the  Spanish  Ministry  of  Education  and  Vocational Training  (MEFP)  through  the  Beatriz  Galindo program  2018  (BEAGAL18/00203) and Spanish    Ministry    MINECO (FIDEUA   PID2019-106901GBI00/10.13039/501100011033). 
\end{acknowledgments}


\appendix

\section{Density of states derivation}

To derive the density of states, we first calculate the total number of states (with any symmetry or parity) and energy less than $E$:
\begin{eqnarray}\label{eq:nos}
    \mathcal{N}(E) &=&  \frac{1}{2^{N}}\frac{\pi^{N/2}}{\Gamma(N/2+ 1)} E^{N/2}\\
    && + O\left[N,W,\tau,\gamma\right](E^{(N-1)/2}).\nonumber
\end{eqnarray}
The density of states in Eq.~(\ref{eq:dos}) is the derivative of this with respect to energy.

To establish this result (\ref{eq:nos}), first consider the simplest limiting case $H(N,W,0,0)$. There is a solution of $H(N,W,0,0)$ for every set of non-negative integers ${\bf n} = \{n_1,\ldots,n_N\}$ with energy $E_\mathbf{n}  = \sum_{i=1}^N n_i^2$. The space of solutions therefore is a (hyper)cubic lattice in the all-positive `quadrant' (really $2^N$-rant) of $\mathbb{R}^N$. Since each state takes up a unit volume in this quantum number space, to find the number of states $\mathcal{N}_N(E)$ with energy less than $E$, one takes the volume of an $N$-ball with radius $r=\sqrt{E}$ and divides by $2^N$ to account for the all-positive condition, giving the leading term in Eq.~(\ref{eq:nos}). The first correction term comes from the $N$-sphere boundary of the $N$-ball, which has one dimension lower.

\begin{figure}
    \centering
    \includegraphics[trim={6cm 4cm 7cm 4cm},clip,width=.9\columnwidth]{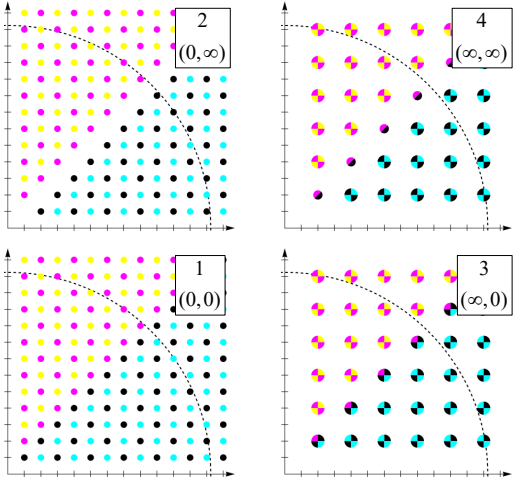}
    \caption{These four diagrams depict the spectrum for the four solvable cases of $N=2$ particle and $W=2$ wells; c.f.~Fig.~\ref{fig:modelspace} and Sec.~\ref{sec:solvable}. Model 1: no interactions or barriers; model 2: infinite interactions no barriers; model 3 no interactions, infinite barriers; model 4 infinite interactions and barriers. Each dot represents an energy eigenstate with energy $E= n_1^2 + n_2^2$ given by the sum of the squares of the integer coordinates $(n_1,n_2)$ of the point. States with $n_1 \geq n_2$ represent the symmetrized states with positive parity in black and with with negative parity  in cyan. States with $n_1 < n_2$ represented by magenta and yellow dots are antisymmetric and have negative and positive parity, respectively. In models 1 and 2, all states are two-fold degenerate except in model 1 (no barriers, no interactions) when $n_1 = n_2$. Additionally, models 3 and 4 have additional two-fold degeneracies (in model 4, states along the diagonal represented by split dots with twice the area), four-fold (in model 4, represented by quartered dots along the diagonal) and eight-fold (in models 3 and 4, represented by pairs of quartered dots with exchanged integer quantum numbers). The dashed quarter-circle is the boundary $E= 150$ with radius $\sqrt{150}$ and is included to aid visualization of the spectral flow.}
    \label{fig:dots}
\end{figure}

The spectrum of $H(2,2,0,0)$ is depicted as model 1 in Fig.~\ref{fig:dots}. In this simple case, states with energy less than $E$ lie within the quarter circle with radius $r = \sqrt{E}$. That quarter disk therefore has area $\pi E/4$, agreeing with (\ref{eq:nos}).

Parallel arguments give the same leading terms for the other three solvable corner models. For example, in the limit of no barriers and infinite interactions $H(N,W,0,\infty)$, we follow the construction of Girardeau and find that there are $N!$ states for every strictly increasing set of integers $\{ {\bf n} | n_1 < n_2 < \cdots n_N \}$; c.f.\ \cite{harshman_one-dimensional_2016}. Therefore, infinite interactions exclude cases when two or more quantum numbers are the same. This is depicted for  $H(2,2,0,\infty)$ in Fig.~\ref{fig:dots}, where the spectrum is missing the diagonal states with $n_1 = n_2$. For $N=2$, the number of states missing from the area estimation in Eq.~(\ref{eq:nos}) is therefore proportional to the length of this boundary $r = \sqrt{E}$. More generally for $N$ particles, the geometrical structures in the quantum number `quadrant' corresponding to all numbers different remains $N$-dimensional even in the presence of infinite interactions. However, the structures with two quantum numbers equal correspond to interior boundaries of the quadrant and have dimension $N-1$. Therefore corrections to account for the `missing states' appear at the subleading order $r^{N-1} = E^{(N-1)/2}$. Further corrections to the number of states appear at next-to-subleading order $E^{(N-2)/2}$ when either two pairs of quantum numbers are the same or three quantum numbers are the same.

Note that in the example with $N=2$ in Fig.~\ref{fig:dots}, the Tonks-Girardeau map shifts the symmetric states $(n_1, n_2)$ in model 1 to $(n_1 +1 , n_2)$ in model 2. Since an integrable model connects these two limiting cases, this establishes a one-to-one adiabatic mapping between the two spectra. From this mapping the number of level crossings that occur as $\gamma$ is tuned from $0$ to $\infty$ can be explicitly calculated without actually solving for the spectrum on the integrable model that links these two cases. 

Similarly, for the case $H(N,W,\infty,0)$ of infinite barriers and no interactions, there are $W^N$ solutions for every set of non-negative integers ${\bf n}$ where all integers $n_i$ are multiples of the number of wells $W$. This increases the volume associated with each set of quantum numbers from $1$ to $W^N$ and that factor of $1/W^N$ exactly cancels the degeneracy factor $W^N$ giving the same leading term. This is depicted for the simplest case of $N=2$ and $W=2$ in model 3 of Fig.~\ref{fig:dots}. As before, we do expect the coefficient on the subleading term to depend on the intricate combinatorics of putting $N$ identical particles in $W$ wells. A parallel argument holds for the fourth solvable model $H(N,W,\infty,\infty)$. 

Note again that these four models with exact solutions are connected by integral models and exact spectral maps can be constructed for all of these cases. Because these four extreme cases all have the same leading term in the density of states that depends only on $N$, we assume that all the models that lie in this region of  parameter space have the same property, and this  is further  supported by the phase space argument presented in Sec.~\ref{sec:density}.

When we elect to consider only one symmetry sector, e.g.~bosons with positive parity, that reduces the number of states (\ref{eq:nos}) by a factor of $1/2$ for parity and $1/N!$ for symmetrization at leading order. At subleading order, the correction coefficient depends on the combinatorics of $N$ particles in $W$ wells and on the parameters $(\tau,\gamma)$. For example, for the simplest case of $N=2$ and $W=2$ depicted, we see the importance of states along the diagonal $n_1=n_2$, which would result in corrections of order $\sqrt{E}$ for the length of that diagonal.

As a final comment, if the leading term of the density of states is independent of the parameters $(\tau, \gamma)$, then the subleading term contains information about the density of level crossings for the integrable models which presumably becomes level repulsions in the non-integrable parameter region. Future work will investigate this connection between spectral flow, level crossings, and energy level statistics.

\bibliographystyle{plainnat}
\bibliography{biblio}

\end{document}